\documentclass[12pt]{article}

\newcommand{\sect}[1]{\setcounter{equation}{0}\section{#1}}

\newcommand{\eq}{\begin{equation}}
\newcommand{\eqa}{\begin{eqnarray}}
\newcommand{\en}{\end{equation}}
\newcommand{\ena}{\end{eqnarray}}
\newcommand{\enn}{\nonumber \end{equation}}

\def\sk{\vskip .4cm}
\def\noi{\noindent}
\def\om{\omega}
\def\al{\alpha}

\def\ga{\gamma}
\def\Ga{\Gamma}

\let \part\partial

\def\unmezzo{{1 \over 2}}
\def\epsi{\varepsilon}
\def\we{\wedge}

\def\de{\delta}

\def\part{\partial}

\def\sk{\vskip .4cm}

\def\noi{\noindent}

\def\X0{X^0}

\def\om{\omega}

\def\al{\alpha}
\def\ga{\gamma}

\def\unmezzo{{1 \over 2}}
\def\epsi{\varepsilon}

\def\we{\wedge}

\def\de{\delta}

\def\Lcal{{\cal L}}

\def\square{{\,\lower0.9pt\vbox{\hrule \hbox{\vrule height 0.2 cm
\hskip 0.2 cm \vrule height 0.2 cm}\hrule}\,}}

\def\Lcal{{\cal L}}

\def\Phi{\phi}
\def\rf{{\rm f}}

\def\fbaralup{\overline{\rf}^\alpha}
\def\fbaraldo{\overline{\rf}_\alpha}

\def\westar{\we_\star}

\def\omtilde{\tilde \om}
\def\Vtilde{\tilde V}
\def\Ttilde{\tilde T}

\def\rtilde{\tilde r}
\def\epsitilde{\tilde \epsi}

\def\psibar{\bar \psi}
\def\Om{\Omega}

\begin{document}

\begin{titlepage}
\rightline{DISTA-UPO/09}
\rightline{February 2009}
\vskip 2em
\begin{center}{\bf NONCOMMUTATIVE D=4 GRAVITY COUPLED TO FERMIONS}
\\[3em]
{\bf Paolo Aschieri}${}^{1,2}$ and {\bf Leonardo Castellani}${}^{2}$ \\ [3em] {\sl ${}^{1}$
Centro ``Enrico Fermi", Compendio Viminale, 00184 Roma, Italy \\
[1em] ${}^{2}$ Dipartimento di Scienze e Tecnologie avanzate and
\\ INFN Gruppo collegato di Alessandria,\\Universit\`a del Piemonte Orientale,\\ Via Bellini 25/G 15100
Alessandria, Italy}\\ [1.5em]
\end{center}

\begin{abstract}
We present a noncommutative extension of Einstein-Hilbert gravity
in the context of twist-deformed space-time, with a $\star$-product associated to
a quite general triangular Drinfeld twist. In particular the $\star$-product can be chosen to be the usual
 Groenewald-Moyal product. 
The action is geometric, invariant
under diffeomorphisms and centrally extended Lorentz
$\star$-gauge
transformations. In the commutative limit it reduces to ordinary
gravity, with local Lorentz invariance and usual real vielbein.
This we achieve by imposing a charge conjugation condition on the
noncommutative vielbein. The theory is coupled to fermions, by
adding  the analog of the Dirac action in curved space. A
noncommutative Majorana condition can be imposed, consistent with
the $\star$-gauge transformations. Finally, we discuss the
noncommutative version of the Mac-Dowell Mansouri action,
quadratic in curvatures.

\end{abstract}

\vskip 6cm \noi \hrule \vskip.2cm \noi {\small
leonardo.castellani@mfn.unipmn.it\\ aschieri@to.infn.it}

\end{titlepage}

\newpage
\setcounter{page}{1}

\sect{Introduction}

Field theories on noncommutative twisted spaces have been the
object of active and recent research: they can be considered as
field theories on ordinary spacetime where the product of fields
is deformed into a twisted, noncommutative and associative
$\star$-product. This product generates infinitely many
derivatives on the fields and introduces a dimensionful
noncommutativity parameter $\theta$. Usually the first step is to
take the classical theory and deform it by replacing ordinary
products by $\star$-products. This is a way to deform a theory by
introducing an infinite number of new interactions and higher
derivative terms.  Some noncommutative deformations of scalar field
theories exhibit nontrivial symmetries \cite{Langmann}  and
provide new renormalizable models \cite{wulkenhaar}, others
lead to new nontrivial integrable systems \cite{GrosseSteinacker}. 
Noncommutative gauge theories have been intensively studied: they naturally
arise under $T$-duality \cite{Tduality}, and also describe a low energy sector
of D-branes physics \cite{SW}. 
The renormalizability of these theories is still problematic. 
Some partial results can be found in ref.s \cite{NCrenorm}.

Noncommutative gravity theories have been constructed in the past
in the context of particular quantum groups \cite{Castellaniqgrav}
and more recently in the twisted noncommutative geometry setting
\cite{Chamseddine,Wessgroup,Estradaetal}. In the second order formalism of ref.
\cite{Wessgroup} the deformed theory is invariant under
diffeomorphisms, but no gauge invariance on the tangent space
(generalizing local Lorentz symmetry) is considered, and
therefore coupling to fermions has not been discussed. In ref. 
\cite{Chamseddine} the noncommutative gravity action has a local
$GL(2,C)$ invariance acting on tangent indices, but reduces in the
commutative limit to gravity with a complex vielbein.
Other attempts to formulate noncommutative deformations
of gravity in the first order formalism can be found in  \cite{ARBA}.

 In this paper, by using the tools of twisted
 differential geometry, we construct a geometric theory of 
 noncommutative gravity.
 The Lagangian is seen to be a globally defined 4-form, hence
 invariant under diffeomorphisms as well as $\star$-diffeomorphisms.
Since these latter  do not change the $\star$-product, they are a
symmetry of the theory.  The action is also invariant under a $GL(2,C)$ $\star$-gauge
 symmetry\footnote{In this paper noncommutative gauge symmetries are
   called for short $\star$-gauge symmetries and should not be
   confused with the twisted gauge symmetries discussed in \cite{twgauge}.}
that reduces to ordinary local Lorentz symmetry in the
 commutative limit. This we achieve by $\star$-extending the
 first order formalism of gravity coupled to Dirac fermions,
formulated in a convenient index-free form.
 For the bosonic part our treatment does not differ much from the approach
 of Chamseddine. However we find a charge conjugation condition on
 the noncommutative vierbein field,
consistent with the $\star$-gauge variations,
 that ensures the usual commutative limit. Thus we do not have
 to cope with an extra vierbein in the $\theta \rightarrow 0$
 limit, as in \cite{Chamseddine}.
 The charge conjugation condition involves also the $\theta$
 dependence of the fields. These we can imagine expanded in powers of
 $\theta$, and in principle this picture introduces infinitely many
 fields, one for each power of $\theta$.
 If we wish, we can use the Seiberg-Witten map to express all fields in terms of the
 classical one, thereby ending up with a finite number of fields.

 We then discuss
 a noncommutative Majorana condition, that allows coupling
 of noncommutative gravity to Majorana fermions.
 The coupling of first order gravity to a Rarita-Schwinger fermion (gravitino),
 a noncommutative generalization of $D=4$,  $N=1$ supergravity, 
 is discussed in a companion paper \cite{AC2}.

 The quantum treatment of the resulting higher-derivative theory
 is still virgin territory: in this case, at least, we cannot do worse
 than in the commutative limit, where the theory is not
 renormalizable and not finite. We do not expect the infinitely
 many derivatives to conspire with the infinities present in the
 commutative theory, and miraculously cancel its divergences.
 In fact the Drinfeld twist corresponds to a braiding matrix
 with unit square, and this leads to a differential geometry with usual derivatives, i.e.
infinitesimal difference operators. On the other hand, we know
that on quasi-triangular quantum groups derivatives become finite
difference operators:  then one can indeed expect that
 theories constructed with the tools of quasi-triangular differential
 geometry may be regulated by noncommutativity. In this perspective
 the present study of twisted gravity may be seen as a preparatory
 step.

 The paper proceeds as follows. In Section 2 we recast the usual
 first order gravity coupled to a fermion field in an index-free
 form, convenient for its noncommutative extension. Section 3
 recalls some basic tools of twisted noncommutative geometry, and
 in Section 4 and 5 we present the action and the invariances of the
 noncommutative theory. Section 6 deals with the noncommutative
 version of MacDowell-Mansouri quadratic gravity. An Appendix on
 gamma matrices summarizes our conventions.

\sect{First order gravity coupled to fermions }

\subsection{Action}

 The usual action of first-order gravity coupled to fermions can be recast
in an index-free form, convenient for generalization to the
non-commutative case:

\eq S =  \int Tr \left(i R \we V \we V \ga_5- [(D\psi) \psibar -
\psi D\psibar] \we V \we V \we V \ga_5 \right) \label{action1}
\en

\noi The fundamental fields are the 1-forms $\Om$  (spin
connection), $V$ (vielbein) and the fermionic 0-form $\psi$ (spin
1/2 field). The curvature 2-form $R$ and the exterior covariant
derivative on $\psi$ are defined by
 \eq
  R= d\Om - \Om \we
\Om, ~~~~~ D\psi = d\psi - \Om \psi
\en
\noi with \eq \Om = {1 \over 4} \om^{ab} \ga_{ab}, ~~~~~V = V^a
\ga_a ~~~~~~
\en
\noi and thus are $4 \times 4$ matrices in the spinor representation.
 See Appendix A for $D=4$ gamma matrix conventions and useful relations.
  The Dirac conjugate is
defined as usual: $\psibar = \psi^\dagger \ga_0$.
 Then also $(D\psi) \psibar$, $\psi D\psibar$ are
matrices in the spinor representation, and the trace
 $Tr$ is taken on this representation.
Using the $D=4$ gamma matrix  identities:
\eq
\ga_{abc} = i \epsi_{abcd} \ga^d \ga_5,~~~~~~~
 Tr (\ga_{ab} \ga_c \ga_d \ga_5) = -4 i \epsi_{abcd}
  \en
 \noi leads to the usual action:
   \eq
   S = \int  R^{ab} \we V^c \we V^d   \epsi_{abcd} + i [ \psibar  \ga^a D\psi -
   (D \psibar)  \ga^a \psi ] \we V^b \we V^c \we V^d  \epsi_{abcd}
   \label{action1comp}
\en
\noi with
\eq
  R \equiv {1\over4} R^{ab} \ga_{ab},~~~R^{ab} = d\om^{ab}  - \om^a_{~c} \we \om^{cb}
\en

\subsection{Invariances}

The action is invariant under local diffeomorphisms
 (it is the integral of a 4-form on a 4-manifold)
  and under the local Lorentz rotations:
\eq
\de_\epsilon V = -[V,\epsilon ] , ~~~\de_\epsilon \Om = d\epsilon - [\Om,\epsilon],~~~~
\de_\epsilon \psi = \epsilon \psi, ~~~\de_\epsilon \psibar = -\psibar \epsilon
\en
\noi with
\eq
 \epsilon = {1\over 4} \epsilon^{ab} \ga_{ab}
  \en
  The invariance can be directly checked on the action (\ref{action1}) noting that
   \eq
  \de_\epsilon  R = - [{ R},\epsilon ]  ~~~\de_\epsilon D\psi = \epsilon D\psi,~~~\de_\epsilon ((D\psi) \psibar)
  = -[ (D\psi) \psibar, \epsilon], ~~~\de_\epsilon (\psi D\psibar) =
      -[\psi D \psibar, \epsilon],
   \en
   \noi using the cyclicity of the trace $Tr$ (on spinor indices) and the fact that $\epsilon$ commutes with
   $\ga_5$. The Lorentz rotations close on the Lie algebra:
   \eq
   [\de_{\epsilon_1},\de_{\epsilon_2}] = -\de_{[\epsilon_1,\epsilon_2]}
   \en

\subsection{Hermiticity and charge conjugation}

Since the vielbein $V^a$ and the spin connection $\om^{ab}$ are
real fields, the following conditions hold:

\eqa
 & & \ga_0 V \ga_0 = V^\dagger,~~~-\ga_0 \Omega \ga_0 =
 \Omega^\dagger, \\
 & &
  \ga_0 [(D\psi) \psibar] \ga_0 =  [\psi D \psibar]^\dagger,~~~
   \ga_0 [\psi D \psibar] \ga_0 =  [(D\psi)  \psibar]^\dagger
 \label{hermconj} \ena

\noi and can be used to check that the action (\ref{action1}) is
real.

 Moreover, if $C$ is the $D=4$ charge conjugation matrix
 (antisymmetric and squaring to $-1$), we have

\eq
 C V C = V^T,~~~C \Omega C = \Omega^T \label{conjVOm}
 \en
 \noi since the matrices $C\ga_a$ and $C \ga_{ab}$ are symmetric.

 Similar relations hold for the gauge parameter $\epsilon= (1/4) \epsi^{ab} \ga_{ab}$:
 \eq
 - \ga_0 \epsilon \ga_0 =  \epsilon^\dagger, ~~~C \epsilon C = \epsilon^T
  \en
  \noi $\epsi^{ab}$ being real.

  The charge conjugation of fermions:
  \eq
  \psi^C \equiv C (\psibar)^T
  \en
  \noi can be extended to the bosonic fields $V$, $\Om$:
   \eq
   V^C \equiv - C V^T C,~~~\Om^C \equiv C \Om^T C
   \en
   \noi Then the relations (\ref{conjVOm}) can be written as:
    \eq
     V^C = -V,~~~\Om^C = \Om
     \en
     and are the analogues of the Majorana condition for the
     fermions:
    \eq
   \psi^C = \psi~~~\rightarrow \psibar = \psi^T C
   \en
    \noi Note also that
   \eq
    (V  \psi)^C = V^C \psi^C
    \en
    In particular, if $\psi$ is a Majorana fermion, $V  \psi$
    is anti-Majorana.
    \sk
    So far we have been treating $\psi$ as a Dirac fermion, and
    therefore reality of the action requires both terms in
    square brackets in the action (\ref{action1}) or
    (\ref{action1comp}).
    If $\psi$ is Majorana, the two terms give the same
    contribution, and only one of them is necessary.

\subsection{Field equations}

 Using the cyclicity of $Tr$ in (\ref{action1}),
 the variation of $V$ , $\Omega$ and $\psibar$ yield respectively the Einstein equation, the torsion equation and the (massless) Dirac equation in index-free form:
 \eqa
 & &Tr \Big(\ga_{a}\ga_5 [i V \we R + i R \we V - X \we V \we V -
  V \we X \we V  -  V \we V \we X ]\Big) =0, \nonumber \\
 & & Tr\Big(\ga_{ab}[i T \we V -i V \we T + \psi \psibar V \we V \we V
   - V \we V \we V \psi \psibar]\Big) =0 \label{torsioneq}
  \\
  & & V \we V \we V \we D\psi - (T \we V \we V - V \we T \we V + V \we V \we T)
  \psi =0 \label{Diraceq}
  \ena
  \noi with
  \eq
 X \equiv  (D \psi)\psibar- \psi D\psibar
  \en
  and where the torsion $T = T^a \ga_a$ is given by:
  \eq
  T \equiv dV -\Om \we V - V \we \Om
   \en
   \noi The torsion equation can be solved, and yields the known
   result:
   \eq
   T^a = 6 i~ \psibar \ga_b \psi~ V^b \we V^a
   \en
  The Dirac equation (\ref{Diraceq}) contains an extra term proportional to the
  torsion: this is due to requiring a real action for gravity
  coupled to Dirac fermions. If one uses the (complex) Dirac
  action
  \eq
  S_{Dirac} = - \int Tr [(D\psi) \psibar \we V \we V \we V \ga_5]
  \label{actionDirac}
  \en
  \noi the torsion terms in the Dirac equation (\ref{Diraceq})
  are not present.

\section{Twist differential geometry: some tools}

The noncommutative deformation of the gravity theories we construct
in the next Sections relies on the existence (in the deformation quantization context, see
for ex \cite{book} ) of an associative $\star$-product between
functions and more generally an associative $\westar$ exterior product between forms, 
satisfying the following properties:
\sk
\noi 
$\bullet~~$ \noi Compatibility with the undeformed exterior differential:
\eq
d(\tau\wedge_\star \tau')=d(\tau)\wedge_\star \tau'=\tau\wedge_\star
d\tau'
\en
$\bullet~~$ Compatibility with the undeformed integral (graded cyclicity property):
        \eq
       \int \tau \westar \tau' =  (-1)^{deg(\tau) deg(\tau')}\int \tau' \westar \tau\label{cycltt'}
       \en
      \noi with $deg(\tau) + deg(\tau')=$D=dimension of the spacetime
      manifold, and where here $\tau$ and $\tau'$ have compact support
      (otherwise stated we require (\ref{cycltt'}) to hold up to
      boundary terms).
\sk
\noi $\bullet~~$ Compatibility with the undeformed complex conjugation:
\eq
       (\tau \westar \tau')^* =   (-1)^{deg(\tau) deg(\tau')} \tau'^* \westar \tau^*
\en
 We describe here a (quite wide) class of twists whose   $\star$-products
 have all these properties.
In this way we have constructed a wide class of noncommutative
deformations of gravity theories. Of course as a particular case we
have the Groenewold-Moyal $\star$-product
\begin{equation}
f\star g = \mu \big{\{} e^{\frac{i}{2}\theta^{\rho\sigma}\partial_\rho \otimes\partial_\sigma}
f\otimes g \big{\}} , \label{MWstar}
\end{equation}
where the map $\mu$  is the usual pointwise
multiplication: $\mu (f \otimes g)= fg$, and $\theta^{\rho\sigma}$ is a constant
antisymmetric matrix.

\subsection{Twist}

\noi Let $\Xi$ be the linear space of smooth vector fields on a smooth manifold $M$, and $U\Xi$ its
universal enveloping algebra. A twist  ${\cal F} \in U\Xi \otimes U\Xi$
defines the associative twisted product
\begin{eqnarray}
f\star g &=& \mu \big{\{} {\cal F}^{-1} f\otimes g \big{\}}
\end{eqnarray}
\noi  where the map $\mu$  is the usual pointwise 
multiplication: $\mu (f \otimes g)= fg$. The product associativity relies on the defining properties of the twist \cite{Wessgroup,book,Aschieri}. 
Using the standard notation
\eq {\cal F}
\equiv \rf^\alpha \otimes \rf_\alpha,   ~~~
  {\cal F}^{-1}
\equiv \overline{\rf}^\alpha \otimes \overline{\rf}_\alpha
\en
 \noi (sum over $\alpha$ understood) where $\rf^\al, \rf_\al, \overline{\rf}^\alpha , \overline{\rf}_\alpha$ are elements of  $U\Xi$, the
 $\star$-product is expressed in terms of ordinary products as:
  \eq
  f \star g =  \overline{\rf}^\alpha (f)  \overline{\rf}_\alpha (g)
   \en
   \noi Many explicit examples of twist are provided by the so-called abelian twists:
\eq
{\cal F}= e^{-\frac{i}{2}\theta^{ab}X_a \otimes X_b} \label{Abeliantwist}
\en
where $\{X_a\}$ is a set of mutually commuting vector fields globally
defined on the manifold\footnote{
We actually need only the twist $\cal F$ to be globally defined, not
necessarily the single vector fields $X_a$.  An explicit example of
this latter kind is given by the  twist 
(\ref{Abeliantwist}), that in an open neighbourhood with
coordinates $t,x,y,z$ is defined by the commuting vector fields 
$X_1=f(x,z){\partial\over \partial x}$, 
$X_2=h(y,z){\partial\over \partial y}$, where $f(x,z)$ is
a function of only the $x$ and $z$ variables and has
compact  support, and similarly $h(y,z)$.  This twist is globally defined on the whole manifold by
  requiring it to be the identity $1\otimes 1$ outside the $\{x^a\}$
  coordinate neighbourhood. The corresponding $\star$-product,
  defined on the whole spacetime manifold, is noncommutative only
  inside this neighbourhood.
}, and $\theta^{ab}$ is a constant
antisymmetric matrix. The corresponding $\star$-product is in general
position dependent because the vector fields $X_a$ are in general
$x$-dependent. In the special case that there exists a
global coordinate system on the manifold we can consider the
vector fields $X_a={\partial \over \partial x^a}$. In this instance we have
the Moyal twist, cf. (\ref{MWstar}):
  \eq
   {\cal F}^{-1}=  e^{\frac{i}{2}\theta^{\rho\sigma}\partial_\rho \otimes\partial_\sigma} \label{Mtwist}
   \en
   \subsection{Deformed exterior product}

   \noi The deformed exterior product between forms is defined as
   \eq
   \tau \westar \tau' \equiv \fbaralup  (\tau) \we \fbaraldo (\tau') \label{defwestar}
       \en
       \noi where $ \fbaralup$ and $\fbaraldo$ act on forms via the Lie derivatives
       ${\cal L}_{ \fbaralup} $,  ${\cal L}_{ \fbaraldo} $
       (Lie derivatives along products $uv \cdots$ of elements of  $\Xi$ are defined
       simply by $ {\cal L}_{uv \cdots} \equiv {\cal L}_u {\cal L}_v \cdots$).
     This product is associative, and in particular satisfies:
   \eq
    \tau \westar h \star \tau' = \tau \star h \westar \tau',~~~h \star (\tau \westar \tau') = (h \star \tau) \westar \tau',~~~
    (\tau \westar \tau') \star h = \tau \westar (\tau' \star h)
    \en
    \noi where $h$ is a $0$-form,  i.e. a function  belonging to $Fun(M)$, the $\star$-product
    between functions and one-forms being just a particular case of  (\ref{defwestar}): 
     \eq
     h \star \tau = \fbaralup (h) \fbaraldo  (\tau), ~~~\tau \star h = \fbaralup (\tau) \fbaraldo (h)
     \en

\subsection{Exterior derivative}
        
         \noi The exterior derivative satisfies the usual (graded) Leibniz rule,
         since it commutes with the Lie derivative:
        \eqa
        & & d (f \star g) = df \star g + f \star dg \\
        & & d(\tau \westar \tau') = d\tau \westar \tau'  + (-1)^{deg(\tau)} ~\tau \westar d\tau'
        \ena

       \subsection{Integration: graded cyclicity}
    
        \noi If we consider an abelian twist (\ref{Abeliantwist})
        given by globally defined commuting vector fields $X_a$,
        then the usual integral is cyclic under the $\star$-exterior
        products of forms, i.e., up to boundary terms,
        \eq
       \int \tau \westar \tau' =  (-1)^{deg(\tau) deg(\tau')}\int \tau' \westar \tau
       \en
      \noi with $deg(\tau) + deg(\tau')=$D=dimension of the spacetime
      manifold. In fact we have 
\eq       \int \tau \westar \tau' =    \int \tau \wedge \tau'=
(-1)^{deg(\tau) deg(\tau')}\int \tau' \wedge \tau= 
(-1)^{deg(\tau) deg(\tau')}\int \tau' \westar \tau
\en
For example at first order in $\theta$,
\eq
\int \tau \westar \tau' =    \int \tau \wedge \tau'-{i\over
 2}\theta^{ab}\int{\cal L}_{X_a}(\tau\wedge {\cal L}_{X_b}\tau') 
=
\int \tau \wedge \tau'-{i\over
 2}\theta^{ab}\int d {i}_{X_a}(\tau\wedge {\cal L}_{X_b}\tau')
\en
where we used the Cartan formula ${\cal L}_{X_a}=di_{X_a}+i_{X_a}d$.
\sk
More generally if the twist $\cal F$ satisfies the condition
$S(\fbaralup)\fbaraldo=1$, 
where the antipode $S$ is defined on vector fields
        as  $S(v)=-v$  and is extended to the whole universal
        enveloping algebra $U\Xi$ linearly and antimultiplicatively,
        $S(uv)=S(v)S(u)$, then a similar argument proves the graded
        cyclicity of the integral\footnote{Proof: using Sweedler's coproduct notation
          (cf. \cite{Wessgroup})) we have 
\eqa\tau\wedge_\star \tau'&=&   \fbaralup  (\tau) \we \fbaraldo (\tau') =
 \fbaralup_1  (\tau \we  S(\fbaralup_2)\fbaraldo (\tau'))=
\tau \we  S(\fbaralup)\fbaraldo (\tau')+ {\fbaralup}'_1  (\tau \we
S({\fbaralup}'_2)\fbaraldo (\tau'))\nonumber\\
&=&\tau \we  \tau'+ {\rm total~derivative}\nonumber
\ena
In the last equality we observed that each ${\fbaralup}'_1$ contains at
least one vector field.  Thus use of Cartan's formula implies that
the second addend is a total derivative.}.

        \subsection{Complex conjugation}
    \sk
        \noi If we choose real fields $X_a$ in the definition of the
        twist (\ref{Abeliantwist}),  it is immediate to verify that:
        \eq
        (f \star g)^* = g^* \star f^*\label{starfg*}
        \en
        \eq
        (\tau \westar \tau')^* =   (-1)^{deg(\tau) deg(\tau')} \tau'^* \westar \tau^*\label{startt*}
        \en
        since sending $i$ into $-i$ in the twist (\ref{Mtwist}) amounts to send $\theta^{ab}$ into
        $-\theta^{ab} = \theta^{ba}$, i.e. to exchange the
        order of the factors in the $\star$-product.
\sk
 More in general
        we can consider twists $\cal F$ that satisfy the reality condition
        (cf. Section 8 in \cite{Wessgroup} )
        ${\fbaralup}^* \otimes {\fbaraldo}^*=S(\fbaraldo) \otimes S
        (\fbaralup)$. The $\star$-products associated to these
        twists satisfy properties (\ref{starfg*}), (\ref{startt*}).

\section{Noncommutative gravity coupled to fermions}

\subsection{Action and symmetries}

Here we generalize Section 2 to the noncommutative case, mostly by replacing
exterior products by deformed exterior products. Thus the action becomes:

\eq S =  \int Tr \left(i {R}\westar V \westar V \ga_5 -[ (D\psi)
\star \psibar - \psi \star D\psibar] \westar V \westar V \westar V
\ga_5 \right) \label{action1NC}
\en

\noi with
 \eq R= d\Om - \Om \westar \Om, ~~~~~ D\psi = d\psi -
\Om \star \psi
\en

Almost all formulae in Section 2 continue to hold, with
$\star$-products and $\star$-exterior products. However, the
expansion of the fundamental fields on the Dirac basis of gamma
matrices must now include new contributions:
 \eq
  \Om = {1 \over 4} \om^{ab} \ga_{ab} + i \om 1 + \omtilde \ga_5, ~~~~~V = V^a
\ga_a + \Vtilde^a \ga_a \ga_5  ~~~~~~
\en
\noi Similarly for the curvature :
 \eq
 R= {1\over 4} R^{ab} \ga_{ab} + i r 1 + \rtilde \ga_5
  \en
 \noi and for the gauge parameter:
 \eq
 \epsilon = {1\over 4} \epsi^{ab} \ga_{ab} + i \epsi 1 + \epsitilde \ga_5
  \en
  \noi Indeed now the $\star$-gauge variations read:
  \eq
\de_\epsilon V = -V \star \epsilon + \epsilon \star V,
~~~\de_\epsilon \Om = d\epsilon - \Om \star \epsilon+ \epsilon
\star \Om,~~~~ \de_\epsilon \psi = \epsilon \star \psi,
~~~\de_\epsilon \psibar = -\psibar \star \epsilon
\label{stargauge}\en
 \noi and in the variations for $V$ and $\Om$ also anticommutators of gamma matrices appear,
 due to the noncommutativity of the $\star$-product. Since for example the anticommutator
 $\{ \ga_{ab},\ga_{cd} \}$ contains $1$ and $\ga_5$, we see that the corresponding fields
 must be included in the expansion of $\Om$. Similarly, $V$ must contain a $\ga_a \ga_5$ term due
 to $\{ \ga_{ab},\ga_{c} \}$. Finally, the composition law for gauge parameters becomes:
 \eq
   [\de_{\epsilon_1},\de_{\epsilon_2}] = \de_{\epsilon_2 \star \epsilon_1 -
   \epsilon_1 \star \epsilon_2 }
   \en
   \noi so that $\epsilon$ must contain the $1$ and $\ga_5$ terms, since they appear in the
   composite parameter $\epsilon_2 \star \epsilon_1 - \epsilon_1 \star \epsilon_2$.

   The invariance of the noncommutative action (\ref{action1NC}) under the $\star$-variations is
   demonstrated  in exactly the same way as for the commutative case, noting that
   \eq
  \de_\epsilon R = - R \star \epsilon+ \epsilon \star R, ~~~\de_\epsilon D\psi = \epsilon \star D\psi,~~~\de_\epsilon ((D\psi) \star \psibar) = - (D\psi) \star \psibar \star \epsilon + \epsilon \star (D\psi) \star \psibar
   \en
   \noi etc., and using now, besides the cyclicity of the trace $Tr$ and the fact that
   $\epsilon$ still commutes with $\ga_5$, also the graded cyclicity
   of the integral.
   
   The local $\star$-symmetry satisfies the Lie algebra of $GL(2,C)$, and centrally extends 
   the $SO(1,3)$ Lie algebra of the commutative theory.
   
   Finally, the $\star$-action (\ref{action1NC}) is invariant under diffeomorphisms
   generated by the Lie derivative, in the sense that 
   \eq
    \int \Lcal_v ({\rm 4{-}form}) = \int (i_v d + d i_v) ({\rm 4{-}form}) = \int d (i_v ({\rm 4{-}form}))= boundary~ term
    \en
    since $d({\rm 4{-}form}) =0$ on a 4-dimensional manifold\footnote{ In order to show that the integrand is a globally defined 4-form we need
to assume that the vielbein one-form $V^a$ is globally defined (and therefore that the
manifold is parallelizable), the twisted exterior product  being globally defined
(because the twist is globally defined). If this is the case, then due to the local
$GL(2,C)$ $\star$-invariance the action is
independent of the vielbein used. On the other hand, if the vielbein $V^a$
is only locally defined  in open coverings of the manifold, then we
cannot construct a global 4-form, since the local 
$GL(2,C)$ $\star$-invariance holds only under integration.}.
   
   We have constructed a geometric lagrangian where
    the fields are exterior forms and the $\star$-product is given by the
    Lie derivative action of the twist on forms.   The
    twist $\cal F$ in general is not invariant under the diffeomorphism
    ${\cal L}_v$.  However we can consider the $\star$-diffeomorphisms
    of ref. \cite{Wessgroup} (see also \cite{book}, section 8.2.4),
  generated by the $\star$-Lie derivative. This latter 
  acts trivially on the twist $\cal F$ but satisfies a deformed Leibniz
  rule. $\star$-Lie derivatives generate infinitesimal noncommutative
  diffeomorphisms and leave invariant the action and the twist. They are
  noncommutative symmetries of our action. 

Finally in our geometric action no coordinate indices $\mu,\nu$ appear, and this
implies invariance of the action under (undeformed) general coordinate
transformations\footnote{General coordinate transformations are
  diffeomorphisms of an open coordinate neighbourhood  of the manifold, not of the
  whole manifold.}.
Otherwise stated every contravariant tensor index $^\mu$ is contracted with the
corresponding covariant tensor index $_\mu$, for example
$X_a=X^\mu_a\partial_\mu$ and $V^a=V^a_\mu dx^\mu$.

\subsection{Field equations}

  \noi  Using the cyclicity of $Tr$ and the graded cyclicity of the integral  in (\ref{action1NC}),
 the variation of $V$ , $\Omega$ and $\psibar$ yield respectively the noncommutative Einstein equation,  torsion equation and Dirac equation in index-free form:
 \eqa
  Tr[\Ga_{a,a5} (iV \westar R + i R \westar V - X \westar V \westar V -
  V \westar X \westar V  - V \westar V \westar  X) ] = 0 \nonumber
  \ena
  \eqa
  Tr[\Ga_{ab,1,5}( iT \westar V - iV \westar T + \psi \star \psibar \star V \westar V \westar V-
  V \westar V \westar V \star \psi \star \psibar)] =0 \nonumber \\ \label{torsioneqNC}
  \ena
  \eqa
  V \westar V \westar V \westar  D\psi  - (T \westar V \westar V - V \westar T \westar V + V \westar V      
   \westar T) \star \psi =0 \nonumber
  \ena
  \noi where   $\Ga_{a,a5}$  indicates $\ga_{a}$ and $\ga_a \ga_5$
  (thus there are two distinct equations) and likewise for
 $\Ga_{ab,1,5}$ (three equations corresponding to $\ga_{ab}$, $1$ and
 $\ga_5$). The noncommutative torsion two-form is defined by:
  \eq
  T \equiv T^a \ga_a + \Ttilde^a \ga_a \ga_5 \equiv  dV -\Om \westar V - V \westar \Om
  \en
 The torsion equation (\ref{torsioneqNC}) can be written as:
   \eq
   [i T \westar V - i V \westar T   + \psi \star \psibar \star V \westar V \westar V
   - V \westar V \westar V \star \psi \star \psibar, \ga_5]_+ =0 \label{torsioneq2}
    \en
    \noi Indeed the anticommutator with $\ga_5$ selects the 
    $\ga_{ab}$, $1$ and $\ga_5$ components. This equation can be solved for
    the torsion:
   \eq
   T = {i \over 2} [\psi \star \psibar \star V \westar V + V \westar \psi \star \psibar \star V +
   V \westar V \star \psi \star \psibar,\ga_5]\ga_5
   \en
   \noi as can be verified by substitution into (\ref{torsioneq2}).
   
 \subsection{$\theta$ - dependent fields}

We can rewrite the Moyal twist as:
 \eq
   {\cal F}^{-1}=  e^{\frac{i}{2}\theta \Theta^{\rho\sigma}\partial_\rho
    \otimes\partial_\sigma} \label{Mtwistalpha}
   \en
 \noi where $\theta$ is a dimensionful parameter (so that
 $\Theta^{\rho\sigma}$ is a numerical matrix).
In the spirit of the Seiberg-Witten map \cite{SW}, the fields and the gauge parameter can be
 considered functions of $x$ and $\theta$. Expanding a field $\phi$ in powers of
 $\theta$:
 \eq
 \phi_\theta (x) = \phi_0 (x) + \theta \phi_1 (x) + \theta^2 \phi_2 (x) + ...,~~~
  \epsi_\theta (x) = \epsi_0 (x) + \theta \epsi_1 (x) + \theta ^2 \epsi_2 (x) + ...
 \en
 \noi introduces an infinite tower of $x$ - dependent fields: a
 finite number of them enters in the action (\ref{action1NC}) at
 each given order in $\theta$. At $0$-th order only the classical fields
 $\phi_0 (x)$ contribute. The gauge variations of all $\phi_i$ are deduced
 by expanding the $\star$-gauge transformations in (\ref{stargauge})
 in powers of $\theta$. Clearly the classical fields $\phi_0$
 transform with the classical gauge variations $\de_\epsilon^0$.

If one feels uncomfortable with these new fields
$\phi_i$,  the Seiberg-Witten map can be used to relate the
higher-order fields to the classical ones in a way consistent with
the $\star$ - gauge transformations $\de_\epsilon$:
 \eq
  \de_\epsilon \phi (\phi_0) = \phi( \de_\epsilon^0 \phi_0)
  \en
 so that the $\star$-deformed theory will contain only the $\phi_0$ fields
 \cite{SW,Jurco}.

All the fields $V^a$, $\Vtilde^a$, $\om^{ab}$, $\om$, and $\omtilde$ contained
in the action (\ref{action1NC}) are then
$\theta$-expanded, and the 0-th order action contains their $\theta \rightarrow 0$ limit.

\subsection{Hermiticity and charge conjugation}

Hermiticity conditions can be imposed on $V$, $\Om$ and the gauge parameter $\epsilon$:
\eq
 \ga_0 V \ga_0 = V^\dagger,~~~ -\ga_0 \Omega \ga_0 =
 \Omega^\dagger,~~~ -\ga_0 \epsilon \ga_0 =
 \epsilon^\dagger \label{hermconjNC}
 \en
 \noi Moreover it is easy to verify the analogues of conditions
 (\ref{hermconj}):
 \eq
 \ga_0 [(D\psi) \star \psibar] \ga_0 = [\psi \star D\psibar]^\dagger,~~ \ga_0 [\psi \star D\psibar] \ga_0 
 = [D\psi \star \psibar]^\dagger
\en
\noi These hermiticity conditions are consistent with the gauge variations, as in the
commutative case, and can be used to check that the action (\ref{action1NC}) is
real. On the component fields $V^a$, $\Vtilde^a$, $\om^{ab}$, $\om$, and $\omtilde$, and
on the component gauge parameters $\epsi^{ab}$, $\epsi$, and $\epsitilde$ the hermiticity conditions (\ref{hermconjNC}) imply that they are real fields.

 The charge conjugation relations (\ref{conjVOm}), however, cannot be exported
 to the noncommutative case as they are. Indeed they would imply the vanishing of the
 component fields $\Vtilde^a$, $\om$, and $\omtilde$ (whose presence is necessary in the noncommutative case) and anyhow would not be consistent with the $\star$-gauge variations.

 An essential modification is needed, and makes
 use of the $\theta$ dependence of the noncommutative fields:
\eq
 C V_\theta (x) C = V_{-\theta} (x)^T,~~~C \Omega_\theta (x) C = \Omega_{-\theta} (x)^T,~~~
 C \epsi_\theta (x) C = \epsi_{-\theta} (x)^T
 \en
These conditions can be checked to be consistent with the
$\star$-gauge transformations. For example $ C V_\theta (x)^T C$ can
be shown to transform in the same way as $V_{-\theta} (x)$:
\eqa
\de_\epsilon (C  V_\theta^T C) &= & C (\de_\epsilon V_\theta)^T C = 
C (- \epsilon_\theta^T \star_{-\theta} V_\theta^T
+ V_\theta^T \star_{-\theta} \epsilon_\theta^T )C= \nonumber \\ & =&\epsilon_{-\theta} \star_{-\theta} 
V_{-\theta}
- V_{-\theta} \star_{-\theta} \epsilon_{-\theta}= \de_\epsilon V_{-\theta}
\ena
 where we have used $C^2 = -1$ and the fact that the transposition of a $\star$-product
 of matrix-valued fields interchanges the order of the matrices but not of the $\star$-multiplied
 fields. To interchange both it is necessary to use the "reflected" $\star_{-\theta}$ product obtained
 by changing the sign of $\theta$, since
 \eq
  f \star_\theta g = g \star_{-\theta} f
  \en
  \noi for any two functions $f,g$.

\noi For the component fields and  gauge parameters the charge
conjugation conditions imply:
 \eqa & & V^a_\theta=V^a_{-\theta}, ~~~
\om^{ab}_\theta = \om^{ab}_{-\theta} \\ & & \Vtilde^a_\theta
=-\Vtilde^a_{-\theta}, ~~~ \om^{}_\theta=- \om^{}_{-\theta} ,~~~\omtilde^{}_\theta=
- \omtilde^{}_{-\theta}, \label{cconjonfields}
 \ena
 \noi Similarly
for the gauge parameters:
 \eqa & & \epsi^{ab}_\theta= \epsi^{ab}_{-\theta}
 \\ & &  \epsi^{}_\theta =- \epsi^{}_{-\theta} ,~~~\epsitilde^{}_\theta=-
\epsitilde^{}_{-\theta} \label{cconjonparam}
 \ena

Finally, let us consider the charge conjugate spinor:
 \eq
 \psi^C \equiv C (\psibar)^T 
 \en
 \noi It transforms under $\star$-gauge variations as:
  \eq
   \de_\epsilon \psi^C = C (\de_\epsilon \psibar)^T = C (-\psibar
   \star \epsilon)^T = C (- \epsilon^T \star_{-\theta} \psi^*)=C
   \epsilon^T C \star_{-\theta} C \psi^* = \epsilon^{}_{-\theta} \star_{-\theta}
   \psi^C
   \en
 \noi i.e. it transforms in the same way as $\psi_{-\theta}$. Then
 we can impose the noncommutative Majorana condition:
  \eq
   \psi^C_\theta= \psi^{}_{-\theta}~~ \Rightarrow ~~\psi^\dagger_\theta \ga_0
   = \psi^T_{-\theta}C
   \en

\subsection{Commutative limit $\theta \rightarrow 0$}

In the commutative limit the action reduces to the usual action of gravity coupled to fermions
of eq. (\ref{action1}). Indeed in virtue of the charge conjugation conditions on $V$ and $\Om$,
the component fields  $\Vtilde^a$, $\om$, and $\omtilde$ all vanish in the limit $\theta \rightarrow 0$
(see the second line of (\ref{cconjonfields})), and only the classical spin connection
$\om^{ab}$, vierbein $V^a$ and Dirac fermion $\psi$ survive. Similarly the gauge parameters
$\epsi$, and $\epsitilde$ vanish in the commutative limit.

\section{Component analysis}

We give here the action (\ref{action1NC}) in terms of the
component fields $V^a$, $\om^{ab}$, $\Vtilde^a$, $\om$, and
$\omtilde$, and the gauge variations of these fields.

\subsection{Action for the component fields}

\eqa
 S = & &  \int R^{ab} \westar (V^c \westar V^d - \Vtilde^c \westar \Vtilde^d)
 \epsilon_{abcd}\nonumber \\
 & &  + 2i~ R^{ab} \westar (-V_a \westar  \Vtilde_b + \Vtilde_a \westar
 V_b)\nonumber \\
 & &+ 4i~r \westar(V^a \westar \Vtilde_a - \Vtilde^a \westar
 V_a)\nonumber \\
 & & + 4i~ \rtilde \westar (V^a \westar V_a -\Vtilde^a \westar
 \Vtilde_a)\nonumber \\
 & & + Tr [(D\psi \star \psibar - \psi \star D \psibar)\ga^d] 
 \westar
 [i \epsilon_{abcd}(V^a \westar V^b \westar V^c \nonumber  \\
 & & ~~~- V^a \westar
 \Vtilde^b \westar \Vtilde^c + \Vtilde^a \westar V^b \westar \Vtilde^c -
  \Vtilde^a \westar
 \Vtilde^b \westar V^c) \nonumber \\
  & & ~~~ + V^a \westar V_a \westar \Vtilde_d-V^a \westar
  \Vtilde_a \westar V_d + \Vtilde^a \westar V_a \westar V_d -
  \Vtilde^a \westar \Vtilde_a \westar \Vtilde_d \nonumber \\
   & & ~~~ + V_d \westar V^a \westar \Vtilde_a-V_d \westar
  \Vtilde^a \westar V_a + \Vtilde_d \westar V^a \westar V_a -
  \Vtilde_d \westar \Vtilde^a \westar \Vtilde_a \nonumber\\
   & & ~~~ - V^a \westar V_d \westar \Vtilde_a+V^a \westar
  \Vtilde_d \westar V_a - \Vtilde^a \westar V_d \westar V_a +
  \Vtilde^a \westar \Vtilde_d \westar \Vtilde_a ] \nonumber\\
   & & + Tr [(D\psi \star \psibar - \psi \star D \psibar)\ga^d \ga_5] \westar
 [i \epsilon_{abcd}(V^a \westar V^b \westar \Vtilde^c \nonumber  \\
 & & ~~~- V^a \westar
 \Vtilde^b \westar V^c + \Vtilde^a \westar V^b \westar V^c -
  \Vtilde^a \westar
 \Vtilde^b \westar \Vtilde^c) \nonumber \\
  & & ~~~ + V^a \westar V_a \westar V_d-V^a \westar
  \Vtilde_a \westar \Vtilde_d + \Vtilde^a \westar V_a \westar \Vtilde_d -
  \Vtilde^a \westar \Vtilde_a \westar V_d \nonumber \\
   & & ~~~ + V_d \westar V^a \westar V_a-V_d \westar
  \Vtilde^a \westar \Vtilde_a + \Vtilde_d \westar V^a \westar \Vtilde_a -
  \Vtilde_d \westar \Vtilde^a \westar V_a \nonumber\\
   & & ~~~ - V^a \westar V_d \westar V_a+V^a \westar
  \Vtilde_d \westar \Vtilde_a - \Vtilde^a \westar V_d \westar \Vtilde_a +
  \Vtilde^a \westar \Vtilde_d \westar V_a ] \nonumber \\
\ena

\noi with
 \eqa
  & & R^{ab}= d \om^{ab} - \unmezzo \om^{a}_{~c} \westar \om^{cb} +
  \unmezzo \om^{b}_{~c} \westar \om^{ca} - {i \over 4} (\om^{ab}+
 \westar \om + \om \westar \om^{ab}) - \nonumber \\
  & &~~~~~~  - {i \over 8}  \epsi^{ab}_{~~cd}( \om^{cd} \westar \omtilde +
   \omtilde \westar \om^{cd}) \\
 & &   r = d\om + {1 \over 8} \om^{ab} \westar \om_{ab} + \om \westar
   \om - \omtilde \westar \omtilde \nonumber \\
    & &  \rtilde = d\omtilde -i (\om \westar \omtilde + \omtilde
    \westar \om) + {i \over 16} \epsi_{abcd} \om^{ab} \westar
    \om^{cd}
     \ena

\subsection{Gauge variations}

\eqa
 & & \de_\epsilon V^a = \unmezzo (\epsi^{a}_{~b} \star V^b + V^b \star
 \epsi^{a}_{~b}) + {i \over 4} \epsi^{a}_{~bcd} (\Vtilde^{b}
 \star \epsi^{cd} -  \epsi^{cd} \star \Vtilde^{b}) \nonumber \\
& &~~~~~~~~ + \epsi \star V^a - V^a \star \epsi - \epsitilde \star
\Vtilde^a - \Vtilde^a \star \epsitilde\\
 & & \de_\epsilon \Vtilde^a = \unmezzo (\epsi^{a}_{~b} \star \Vtilde^b + \Vtilde^b \star
 \epsi^{a}_{~b}) + {i \over 4} \epsi^{a}_{~bcd} (V^{b}
 \star \epsi^{cd} -  \epsi^{cd} \star V^{b}) \nonumber \\
& &~~~~~~~~ + \epsi \star \Vtilde^a - \Vtilde^a \star \epsi -
\epsitilde \star V^a - V^a \star \epsitilde\\
 & & \de_\epsilon \om^{ab} = \unmezzo (\epsi^a_{~c} \star \om^{cb} -\epsi^b_{~c} \star \om^{ca}
   + \om^{cb} \star  \epsi^a_{~c} - \om^{ca} \star \epsi^b_{~c})
   \nonumber \\
   & & ~~~~~~~~ + {1 \over 4} (\epsi^{ab} \star \om - \om \star
   \epsi^{ab}) + {i \over 8} \epsi^{ab}_{~~cd} (\epsi^{cd} \star
   \omtilde - \omtilde \star \epsi^{cd})\\
   & & ~~~~~~~~+ {1\over 4} (\epsi \star \om^{ab} - \om^{ab} \star \epsi)
   + {i \over 8} \epsi^{ab}_{~~cd} (\epsitilde \star
   \om^{cd} - \om^{cd} \star \epsitilde)\\
   & & \de_{\epsilon} \om = {1\over 8} (\om^{ab} \star \epsi_{ab} -
   \epsi_{ab} \star \om^{ab}) + \epsi \star \om - \om \star \epsi
   + \epsitilde \star \omtilde - \omtilde \star \epsitilde\\
      & & \de_{\epsilon} \omtilde = {i \over 16} \epsi_{abcd}
    (\om^{ab} \star \epsi^{cd} - \epsi^{cd} \star \om^{ab}) +
    \epsi \star \omtilde - \omtilde \star \epsi + \epsitilde \star
    \om - \om \star \epsitilde
    \ena

\section{Noncommutative Mac-Dowell Mansouri gravity}

\subsection{Action and symmetries}

As already discussed in  \cite{ChamMM}, the noncommutative generalization of the Mac-Dowell Mansouri action \cite{MDM} reads:
\eq
 S = i \int Tr [R \westar R \ga_5] \label{MDM}
 \en
 \noi with
  \eq
  R = d \Om - \Om \westar \Om
  \en
  \noi and
  \eq
  \Om = {1 \over 4} \om^{ab} \ga_{ab} + i \om 1 + \omtilde \ga_5 +
  i V^a \ga_a + i \Vtilde^a \ga_a \ga_5
   \en
   The $GL(2,C)$ $\star$-gauge variations act as:
   \eq
   \de_\epsilon \Om = d\epsilon - \Om \star \epsilon+ \epsilon \star \Om
     \en
   \noi with
  \eq
   \epsilon = {1\over 4} \epsi^{ab} \ga_{ab} + i \epsi 1 + \epsitilde \ga_5
  \en
  \noi so that
   \eq
  \de_\epsilon R = - R \star \epsilon+ \epsilon \star R
   \en
   \noi The invariance of the action (\ref{MDM}) under $\star$-gauge transformations is easily checked,
    taking into account the transformation of $R$, the cyclicity of the trace $Tr$, the graded
    cyclicity of the integral and the fact that $\epsilon$ still commutes with
    $\ga_5$.

\subsection{Hermiticity and charge conjugation}

Hermiticity conditions can again be imposed on $\Om$ and on the
gauge parameter $\epsilon$:
 \eq
 -\ga_0 \Omega \ga_0 =
 \Omega^\dagger,~~~ -\ga_0 \epsilon \ga_0 =
 \epsilon^\dagger
 \en
\noi These conditions are consistent with the gauge variations,
and can be used to check that the action (\ref{MDM}) is real.
Again the hermiticity conditions imply that the component fields
 $V^a$, $\Vtilde^a$, $\om^{ab}$, $\om$, $\omtilde$, and the
component gauge parameters $\epsi^{ab}$, $\epsi$, $\epsitilde$ are
real.

The charge conjugation conditions are again 
 \eq
C \Omega^{}_\theta (x) C = \Omega(x)_{-\theta}^T,~~~
 C \epsi^{}_\theta (x) C = \epsi(x)_{-\theta}^T
 \en
These conditions are consistent with the $\star$-gauge
transformations.

\noi For the component fields and  gauge parameters the charge
conjugation conditions imply the same relations
(\ref{cconjonfields}), (\ref{cconjonparam}) as in Section 4.

\subsection{Commutative limit $\theta \rightarrow 0$}

In the commutative limit the action reduces to the usual action of
Mac Dowell-Mansouri gravity. Indeed the charge conjugation
conditions on  $\Om$ ensure that the component fields $\Vtilde^a$,
$\om$, and $\omtilde$ all vanish in the limit $\theta \rightarrow 0$,
and only the classical spin connection $\om^{ab}$, vierbein $V^a$
survive. Moreover the gauge parameters $\epsi$ and $\epsitilde$
vanish in the limit because of the charge conjugation condition on
$\epsilon$ , and only the parameter $\epsi^{ab}$ corresponding to
Lorentz symmetry survives.

\section{Conclusions}

We have constructed a geometric noncommutative action of
first-order gravity coupled to fermions,  invariant under $\star$-diffeomorphisms and
$GL(2,C)$ $\star$-gauge
transformations. The commutative limit reproduces the usual action
with no extra fields, and the $\star$-invariance reduces to ordinary Lorentz invariance.
A charge conjugation condition, consistent with the $\star$-symmetries, is imposed
on the noncommutative vielbein and connection, and takes into account their
$\theta$-dependence. This condition allows to recover the usual commutative limit.
Finally, using the same tools of twisted differential geometry, we
find the noncommutative extension of the Mac-Dowell Mansouri action.

\sect{Appendix A : gamma matrices in $D=4$}

We summarize in this Appendix our gamma conventions in $D=4$.

\eqa
& & \eta_{ab} =(1,-1,-1,-1),~~~\{\ga_a,\ga_b\}=2 \eta_{ab},~~~[\ga_a,\ga_b]=2 \ga_{ab}, \\
& & \ga_5 \equiv i \ga_0\ga_1\ga_2\ga_3,~~~\ga_5 \ga_5 = 1,~~~\epsi_{0123} = - \epsi^{0123}=1, \\
& & \ga_a^\dagger = \ga_0 \ga_a \ga_0, ~~~\ga_5^\dagger = \ga_5 \\
& & \ga_a^T = - C \ga_a C^{-1},~~~\ga_5^T = C \ga_5 C^{-1}, ~~~C^2 =-1,~~~C^T =-C
\ena

\subsection{Useful identities}

\eqa
 & &\ga_a\ga_b= \ga_{ab}+\eta_{ab}\\
 & & \ga_{ab} \ga_5 = {i \over 2} \epsilon_{abcd} \ga^{cd}\\
 & &\ga_{ab} \ga_c=\eta_{bc} \ga_a - \eta_{ac} \ga_b -i \epsi_{abcd}\ga_5 \ga^d\\
 & &\ga_c \ga_{ab} = \eta_{ac} \ga_b - \eta_{bc} \ga_a -i \epsi_{abcd}\ga_5 \ga^d\\
 & &\ga_a\ga_b\ga_c= \eta_{ab}\ga_c + \eta_{bc} \ga_a - \eta_{ac} \ga_b -i \epsi_{abcd}\ga_5 \ga^d\\
 & &\ga^{ab} \ga_{cd} = -i \epsi^{ab}_{~~cd}\ga_5 - 4 \de^{[a}_{[c} \ga^{b]}_{~~d]} - 2 \de^{ab}_{cd}
 \ena
 \noi where $\de^{ab}_{cd}
 = \unmezzo (\de^a_c \de^b_d - \de^a_d \de^b_c)$, and index antisymmetrizations in square brackets have weight 1.

 \subsection{Charge conjugation and Majorana condition}

 \eqa
 & &   {\rm Dirac~ conjugate~~} \psibar \equiv \psi^\dagger
 \ga_0\\
 & &  {\rm Charge~ conjugate~spinor~~} \psi^c = C (\psibar)^T  \\
 & & {\rm Majorana~ spinor~~} \psi^c = \psi~~\Rightarrow \psibar =
 \psi^T C
 \ena

 \vfill\eject

\end{document}